\documentclass[conference]{IEEEtran}
\IEEEoverridecommandlockouts
\usepackage{makecell}
\usepackage{cite}
\usepackage{amsmath,amssymb,amsfonts}
\usepackage{algorithmic}
\usepackage{graphicx}
\usepackage{textcomp}
\usepackage{xcolor}
\usepackage{gensymb}
\usepackage{amsmath}
\usepackage{multicol}
\usepackage{float}
\def\BibTeX{{\rm B\kern-.05em{\sc i\kern-.025em b}\kern-.08em
T\kern-.1667em\lower.7ex\hbox{E}\kern-.125emX}}
\begin{document}
\title{Optimization of  Temperature and Relative Humidity  in an Automatic Egg Incubator Using Mamdani Fuzzy Inference System\\
}
\author{\IEEEauthorblockN{Pramit Dutta\textsuperscript{1} and Nafisa Anjum\textsuperscript{2} }
\IEEEauthorblockA{\textit{Department of Electronics and }
{Telecommunication Engineering} \\
\textit{Chittagong University of}
{Engineering and Technology}\\
Chittagong, Bangladesh \\
pramitduttaanik@gmail.com\textsuperscript{1} and
nafisaanjum94@gmail.com\textsuperscript{2}
}
}

\maketitle
\begin{abstract}
Temperature and humidity are two of the rudimentary
factors that must be controlled during egg incubation.
Improper temperature and humidity levels during the incubation
period often result in unwanted conditions. This paper proposes
the design of an efficient Mamdani fuzzy inference system
instead of the widely used Takagi-Sugeno system in this field
for controlling the temperature and humidity levels of an egg
incubator. Though the optimum incubation temperature and
humidity levels used here are that of chicken egg, the proposed
methodology is applicable to other avian species as well. The
input functions have been used here as per estimated values for
safe hatching using Mamdani whereas defuzzification method,
Center of Area (COA), has been applied for output. From the model output, a stabilized heat from temperature level and fan speed to
control the humidity level of an egg incubator can be obtained.
This maximizes the hatching rate of healthy chicks under any
conditions in the field. 
\end{abstract}
\section{Introduction}
The importance of egg incubators in achieving the ultimate goal of poultry industry, that is, to increase the hatching rate and produce healthy hatched birds is undeniable. If temperature is too high or too low during incubation, problems like dead embryos at early stage, underdeveloped or crippled chicks, early hatching and so on may arise. On the other hand, if humidity level is not maintained properly, embryos adhering to shells, dead embryos before pipping, early broiler mortality and many other difficulties arise.
\par Over the past decade, rapid growth has been noticed in consumer
demand in the developing countries for livestock products. Such 
rise of demand is being met by corresponding rise in poultry meat 
and egg production. Comparative case studies have shown that in countries like India, China, Thailand, Brazil and Egypt, the average per capita income plays a vital role in determining the rate of consumption of meat. World egg production has increased by more than 150 percent in the last three decades. This growth has been more noticeable in Asia where production has increased almost fourfold. World poultry meat production shot up from 9 to 122 million tonnes between 1961 and 2017 whereas egg production soared from 15 to 87 million tonnes [1]. By 2025, the global poultry production is expected to experience a rise of 24\% reaching 1,44,682 thousand tonnes approximately [2].
\par In this context, the obvious burning question that needs to be addressed is how effective are the current physical conditions of incubation? Are they effective enough to promote greater hatchability and better quality chicks? Are they effective enough such that under adverse conditions, they can demonstrate better performance in the field and ensure higher survival rate? As a new and future prospect for incubation, an efficient system has been proposed in this paper that stabilises the output of heater and controls the fan speed of an automatic egg incubator. In most of the cases of incubation, temperature is controlled manually whereas humidity level is managed by adding water to the built-in water trays at the bottom of an incubator. The system here takes inputs from a sensor (DHT22) and  artificial intelligence based Mamdani fuzzy inference system designed here compares it with the required optimum values. The output heat is adjusted according to the input temperature and humidifier fan speed is maintained throughout the process for correct humidity. As such embryos do not lose excess water and mortality rate is considerably reduced. 
\section{RELATED WORK}
Many researchers have presented their work on incubators trying to develop more efficient models throughout time. One of them includes the optimization of temperature in incubator using fuzzy inference and IoT [3]. Another paper discusses the design of incubator using STEM approach [4].In addition, a micro-controller based electrical incubator system has been designed [5]. A temperature and humidity controller has been implemented through FPGA [6]. A quail egg incubator has been designed using Arduino micro-controller and IoT [7]. Fuzzy logic control has been used in designing a private home heating system [8].In this paper, comparative study has been conducted on the management of temperature inside egg hatching incubator from implementation of on/off and fuzzy logic controller using a low cost micro-controller board interfaced with the integrated circuit based temperature sensor and LabVIEW 2015 software [9]. Most of the works in this field has been done using the Takagi-Sugeno fuzzy inference system but the method proposed here uses Mamdani in which the output is not maintained constant or linear. This is much more efficient as it takes linguistic inputs of temperature and humidity which do not vary linearly in nature.
\section{METHODOLOGY}
\begin{figure}[htbp]
\includegraphics[width=0.48\textwidth,height=0.3\textheight]{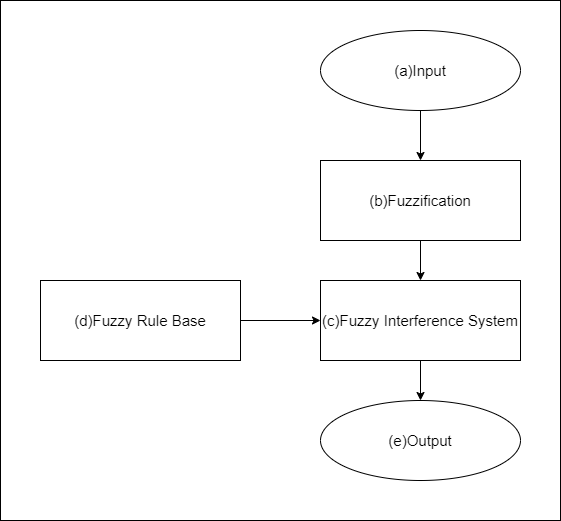}
\caption{Design Procedure}
\label{fig}
\end{figure}
\subsection{Input}The inputs are taken by the sensor DHT22. This sensor has temperature measuring range from -40 to +125 degrees Celsius with $\pm 0.5$ degree accuracy. Moreover, the sensor has much efficient humidity measuring range from 0 to 100\% with 2-5\% accuracy. The operating voltage of DHT22 ranges from 3 to 5 volts while the maximum current used when measuring is 25mA.
\begin{figure}[htbp]
\includegraphics[width=0.4\textwidth,height=0.1\textheight]{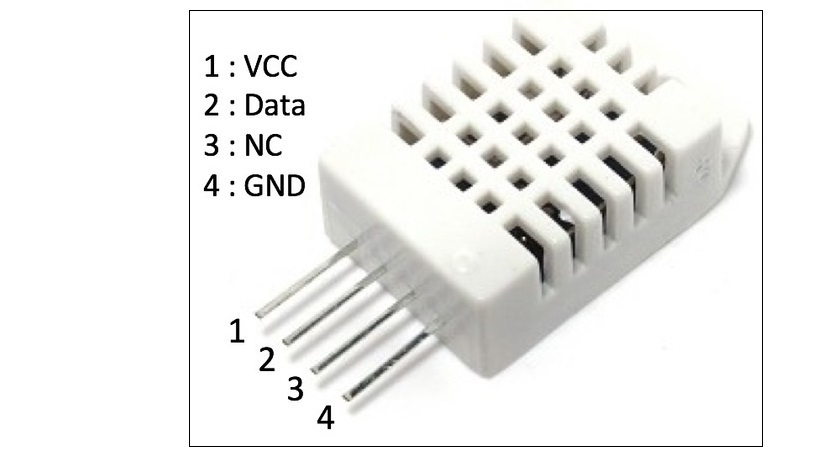}
\caption{DHT22 Sensor}
\label{fig}
\end{figure}
\subsubsection{Temperature}\label{AA}
In a chicken egg incubator it is mandatory to maintain the temperature throughout the entire incubation process. The optimal temperature is 38\degree C(100.5\degree F)but should be kept in a range between 37-39\degree  C.[3]
\subsubsection{Relative Humidity}\label{AA}
Usually for chicken egg, the incubation process is completed in 21 days. The relative humidity should be 50-55\%  in first 17 days and it should be 60-65\% for day 18-21.
\subsection{Fuzzification}\label{AA}In the fuzzification process  all the parameters have been defined as variables according to their value. For the parameters defined, the shape of the membership functions are Triangular and Gaussian function respectively. As per requirement Triangular functions for temperature, humidity and fan speed have been used. Triangular function can be represented as:
\begin{equation*}
\mu\textsubscript{A}(x)=
\begin{cases}
0,\hspace{0.7cm}         x \leq 0
\\
\frac{x-a}{m-a},\hspace{0.2cm}         a<x \leq m
\\
\frac{b-a}{b-m},\hspace{0.2cm}         m<x \leq b
\\
0,\hspace{0.7cm}         x \geq b
\end{cases}
\end{equation*}
Where,\\
a=lower limit;\\
b=upper limit;\\
m=a value\\
The heat change in an incubator should be smooth for stable hatching. So, for heat, Gaussian function has been used. A Gaussian function with central value m and standard deviation k\textgreater0 can be represented as:
\begin{equation*}
\mu\textsubscript{A}(X)=e^{-\frac{(x-m)^2}{2k^2}}
\end{equation*}
For temperature there are three membership functions defined as follows:
\begin{figure}[htbp]
\includegraphics[width=0.48\textwidth,height=0.15\textheight]{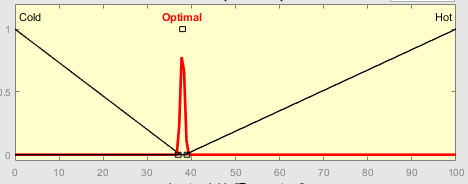}
\caption{Membership Function Plot of Temperature}
\label{fig}
\end{figure}
\par But in case of another input parameter, that is, relative humidity, the condition for membership function has to be defined twice. The first condition is set for first 17 days of incubation process then as per the changes in humidity requirement for day 18-21, changes have been made in the functions as follows:
\begin{figure}[htbp]
\includegraphics[width=0.48\textwidth,height=0.15\textheight]{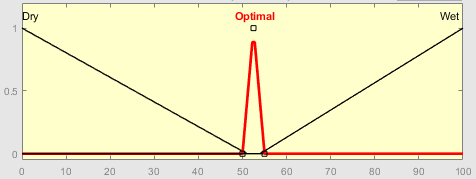}
\caption{Membership Function Plot of Humidity for first 17 days}
\label{fig}
\end{figure}
\par For proper incubation the relative humidity should be increased after 17 days by 10-15 \% which can be defined as:
\begin{figure}[H]
\includegraphics[width=0.48\textwidth,height=0.15\textheight]{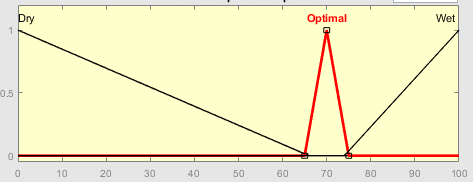}
\caption{Membership Function Plot of Humidity for day 18-21}
\label{fig}
\end{figure}
\par After that, the outputs of incubation process have been taken as heat and fan speed. The membership function of the estimated heat is defined in a range between 1-10. The heat change in an incubator should be smooth for stable hatching. For that reason the Gaussian function has been used as membership function. The design for this is given below:
\begin{figure}[htbp]
\includegraphics[width=0.48\textwidth,height=0.15\textheight]{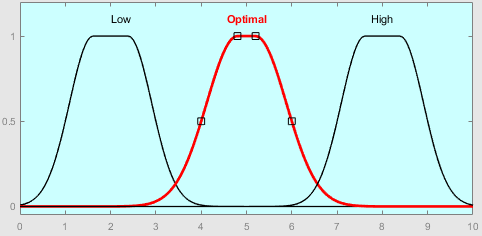}
\caption{Membership Function Plot of Heat}
\label{fig}
\end{figure}
\par The output of fan speed has also been categorized into three membership functions defined in a range of 1-10. The figure can be observed below:
\begin{figure}[htbp]
\includegraphics[width=0.48\textwidth,height=0.15\textheight]{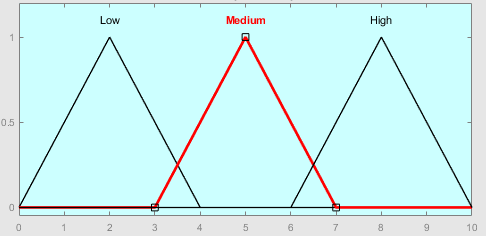}
\caption{Membership Function Plot of Fan Speed}
\label{fig}
\end{figure}
\subsection{Fuzzy Inference system}
There are two methods of FIS. They are:
\par1.Mamdani Fuzzy Inference System;
\par2.Takagi-Sugeno Fuzzy Model (TS Method);
\par In this application, the membership function and relation between them is not linear or constant to gain an optimal temperature and humidity. By considering this characteristic, Mamdani Fuzzy Inference System has been applied. 
\subsection{Fuzzy Rules Base}
\par The rules for the optimization of temperature and humidity in the incubation process have been set for maximum possible conditions that may prevail in the process:
\begin{table}[htbp]
\caption{Rules for fuzzy model}
\begin{center}
\begin{tabular}{|c|c|c|c|c|}
\hline
\textbf{Sl No.}&\textbf{Temperature}&\textbf{\makecell{Relative\\ Humidity}}&\textbf{Heat}&\textbf{Fan Speed} \\
\hline
01.&Cold&None&High&Medium\\
\hline
02.&Optimal&None&Optimal&Medium\\
\hline
03.&Hot&None&Low&Medium\\
\hline
04.&None&Dry&Optimal&High\\
\hline
05.&None&Optimal&Optimal&Medium\\
\hline
06.&None&Wet&Optimal&Low\\
\hline
07.&Cold&Dry&High&High\\
\hline
08.&Cold&Optimal&High&Medium\\
\hline
09.&Cold&Wet&High&Low\\
\hline
10.&Optimal&Dry&Optimal&High\\
\hline
11.&Optimal&Optimal&Optimal&Medium\\
\hline
12.&Optimal&Wet&Optimal&Low\\
\hline
13.&Hot&Dry&Low&High\\
\hline
14.&Hot&Optimal&Low&Medium\\
\hline
15.&Hot&Wet&Low&Low\\
\hline
\end{tabular}
\label{tab1}
\end{center}
\end{table}
\subsection{Output}Here, for defuzzification process, Centroid of Area (COA) Method has been used which can be represented as:\\\\
x\textsuperscript{*}=$\frac{\sum_{i=1}^{k} A\textsubscript{i} \times \bar{x}\textsubscript{i}}{\sum_{i=1}^{k} A\textsubscript{i}}$
\\
where,\\
A\textsubscript{i}=firing area of i\textsuperscript{th} rules;\\
k=total number of rules fired;\\
$\bar{x}\textsubscript{i}$=center of area\\
\subsubsection{Heat}\label{AA}
Relating to the temperature and relative humidity, the heat has been measured here considering estimated value but the actual heat is going to vary depending on the number of eggs and structure of the incubator.
\begin{figure}[htbp]
\includegraphics[width=0.48\textwidth,height=0.25\textheight]{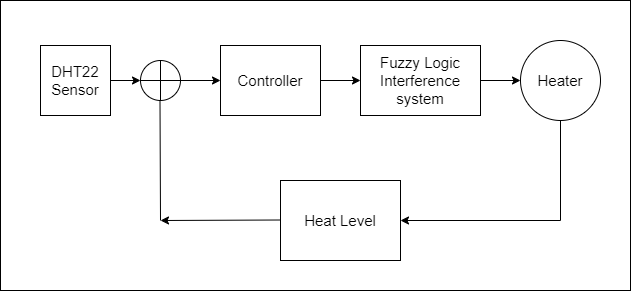}
\caption{Block diagram for heat control}
\label{fig}
\end{figure}
\subsubsection{Fan Speed}\label{AA}The humidifer fan evaporates the water and brings it into the air. this humidifier fan is used with the water channel. When the relative humidity is under the optimal level, the fan speed will be high and vice-versa. 
\begin{figure}[htbp]
\includegraphics[width=0.48\textwidth,height=0.25\textheight]{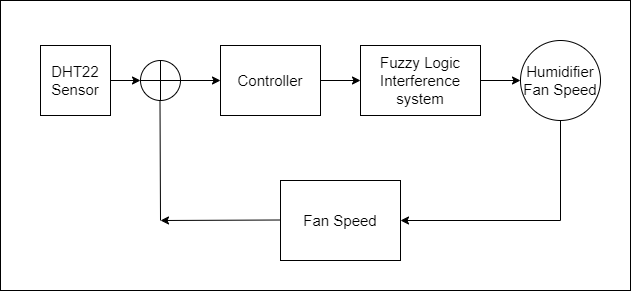}
\caption{Block diagram for fan-speed control}
\label{fig}
\end{figure}
\subsection{Simulation Model}
The simulink model of this control system can be represented as:
\begin{figure}[htbp]
\includegraphics[width=0.48\textwidth,height=0.3\textheight]{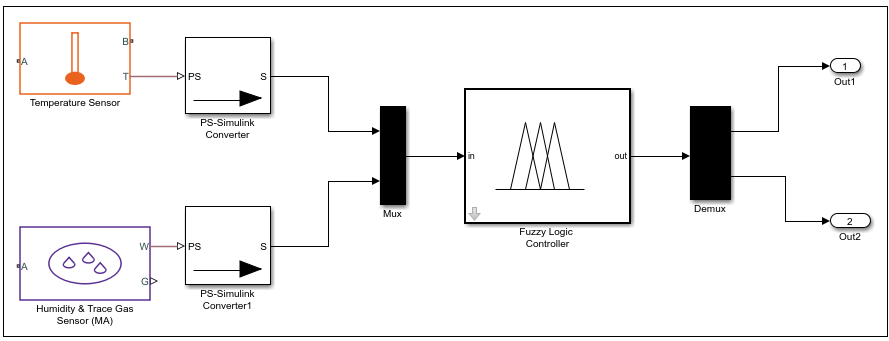}
\caption{Simulation model for the inference system}
\label{fig}
\end{figure}
\subsection{System Diagram}
\par In automatic egg incubators, the temperature and humidity within the incubation chamber is measured by the DHT22 sensor. The fuzzy logic inference system takes these parameters as inputs, compares the data with the stored optimal levels and the generated output controls the heater and humidifier fan as a closed loop feedback system. With the rise of temperature above optimal level, the heat level is adjusted by reducing the heat and vice versa. The humidifier fan speed decreases proportionally to the increase in relative humidity so with decrease in humidity level the ventilation increases whereas when relative humidity is low, the fan speed increases.
\begin{figure}[htbp]
\includegraphics[width=0.48\textwidth,height=0.3\textheight]{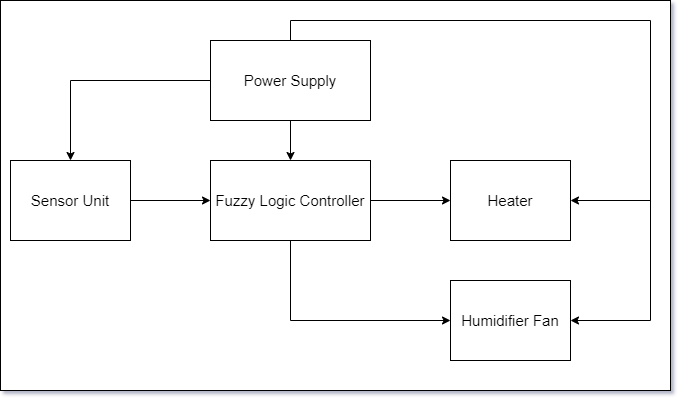}
\caption{Block Diagram}
\label{fig}
\end{figure}
\section{RESULT}
After defuzzification, the result can be analysed through the two outputs. For the first 17 days, it can be observed that when the temperature and humidity were 38\degree C and 53.60\%, the estimate of stabilized heat and fan-speed were found to be 5.00 and 5.00 respectively (considered in a range of 10). 
\begin{table}[H]
\caption{Output for first 17 days}
\begin{center}
\begin{tabular}{|c|c|c|c|}
\hline
\textbf{\makecell{Temperature\\(in \degree C)}}&\textbf{\makecell{Relative\\ Humidity(in \%)}}&\textbf{\makecell{Heat(in\\ scale of 0 to 10)}}&\textbf{\makecell{Fan Spped(in\\ scale of 0 to 10)}} \\
\hline
0.00&34.30&7.02&6.06\\
\hline
24.70&45.20&6.38&5.79\\
\hline
34.30&48.80&5.68&5.37\\
\hline
35.50&51.20&5.33&5.00\\
\hline
38.00&53.60&5.00&5.00\\
\hline
41.60&58.40&4.80&4.40\\
\hline
50.00&17.50&4.25&7.20\\
\hline
58.40&62.00&3.43&3.92\\
\hline
64.50&69.30&3.19&3.65\\
\hline
76.50&75.30&3.09&3.64\\
\hline
\end{tabular}
\label{tab1}
\end{center}
\end{table}
For day 18-21, the criteria for relative humidity was changed as humidity level increased. It was seen that when the temperature and humidity were 38.2\degree C and 64.10\%,the estimate of stabilized heat and fan-speed were found to be 5.00 and 5.00 respectively.
\pagebreak
\begin{table}[H]
\caption{Output for day 18-21}
\begin{center}
\begin{tabular}{|c|c|c|c|}
\hline
\textbf{\makecell{Temperature\\(in \degree C)}}&\textbf{\makecell{Relative\\ Humidity(in \%)}}&\textbf{\makecell{Heat(in scale\\ of 0 to 10)}}&\textbf{\makecell{Fan Spped(in\\ scale of 0 to 10)}} \\
\hline
0.00&34.10&6.98&6.28\\
\hline
22.30&40.40&5.80&6.00\\
\hline
33.10&52.20&5.26&5.47\\
\hline
35.50&59.20&5.06&5.20\\
\hline
38.20&62.60&5.00&5.00\\
\hline
40.40&64.10&4.86&4.17\\
\hline
51.20&19.90&4.25&7.60\\
\hline
57.20&80.10&3.23&3.52\\
\hline
68.10&85.70&3.19&3.25\\
\hline
76.50&87.30&3.06&3.04\\
\hline
\end{tabular}
\label{tab1}
\end{center}
\end{table}
The following graph represents the relation between input and output in the fuzzy logic inference system. By observing the graph, it can be said that when the temperature rises, the heat distribution falls but such changes vary in non-linear fashion. As the humidity within incubation chamber increases, the humidifier fan speed decreases considerably. 
\begin{figure}[H]
\includegraphics[width=0.48\textwidth,height=0.21\textheight]{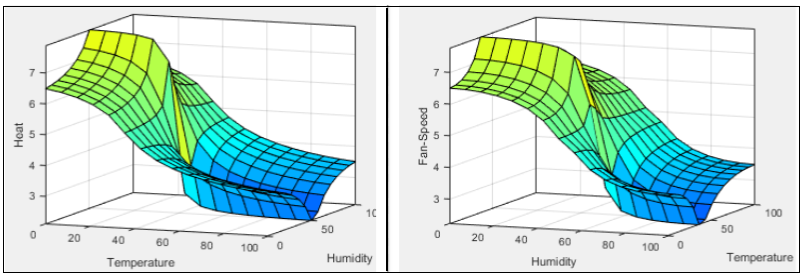}
\caption{Graph representing changes in heat and fan speed for 1-17 days}
\label{fig}
\end{figure}
For day 18-21, the incubation procedure requires additional humidity. During this period, the optimal temperature level remains unchanged and hence, so does the heat. The relation between input and output has been depicted in the following graph:
\begin{figure}[H]
\includegraphics[width=0.48\textwidth,height=0.21\textheight]{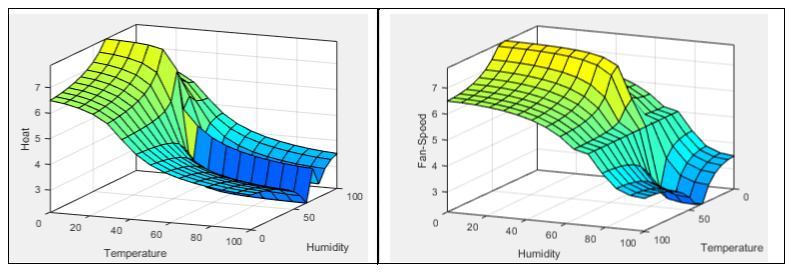}
\caption{Graph representing changes in heat and fan speed for 18-21 days}
\label{fig}
\end{figure}
In an egg incubator, additional 1-2 degree Celsius in temperature can cause early hatching with high mortality rate.On the contrary, reduction of 1-2 degree Celsius in temperature can cause late hatching with reduced hatching rate. Considering such cases, the Mamdani fuzzy inference system proposed in this paper serves as a far better choice than Takagi-Sugeno inference system. This is because Takegi-Sugeno system gives the output of the system as linear or constant in nature. So, the optimization of such systems are not accurate and not up-to the mark. On the other hand, the heat and fan speed changes in this proposed system is not only accurate but also considerably smooth.
\section{CONCLUSION}
In conclusion, it can be said that the Mamdani fuzzy inference system used in designing this model is much suitable for parameter controlling in Multiple Input Multiple Output (MIMO) systems. Though other works have been conducted in this field, most of them have implemented the TS method which does not provide membership functions for the output so the transition of output parameters due to unstable conditions in the field cannot be observed clearly. Temperature and humidity are factors that vary unpredictably and non-linearly in nature and considering such cases, the system presented here produces stabilized output levels by controlling the heater and humidifier fan of an egg incubator. The model proposed here can be applied in the egg incubation process of other avian species as well by changing the crisp values of input parameters. The model can also be modified for use in large scale systems as well. 
\section*{}

\pagebreak
\end{document}